\documentclass[conference]{IEEEtran}

\PassOptionsToPackage{hyphens}{url}

\usepackage[
  hidelinks,
  pdfusetitle,  
]{hyperref}

\usepackage{cite}
\usepackage{amssymb}
\usepackage{booktabs}
\usepackage{fancyhdr}
\usepackage{graphicx}
\usepackage{mathtools}
\usepackage{silence}                    
\usepackage{textcomp}
\usepackage{threeparttable}
\usepackage{xcolor}

\makeatletter
\let\MYcaption\@makecaption
\makeatother
\usepackage[font=footnotesize]{subcaption}
\makeatletter
\let\@makecaption\MYcaption
\makeatother

\usepackage{csquotes}

\usepackage[
  capitalize,
  nameinlink,
]{cleveref}

\MakeOuterQuote{"}

\title{Learned Point Cloud Compression for Classification}

\author{%
  \texorpdfstring{%
    \IEEEauthorblockN{Mateen Ulhaq}
    \IEEEauthorblockA{
      \textit{School of Engineering Science} \\
      \textit{Simon Fraser University}\\
      Burnaby, BC, Canada \\
      mulhaq@sfu.ca%
    }%
    \and%
    \IEEEauthorblockN{Ivan V. Baji\'c}
    \IEEEauthorblockA{
      \textit{School of Engineering Science} \\
      \textit{Simon Fraser University}\\
      Burnaby, BC, Canada \\
      ibajic@ensc.sfu.ca%
    }%
  }{%
    Mateen Ulhaq \and Ivan V. Baji\'c%
  }%
}

\DeclareMathOperator*{\argmax}{arg\,max}

\newcommand{\boldvar}[1]{{\boldsymbol{#1}}}

\newenvironment{subsubfigure}[2][]{%
  \begin{subfigure}[#1]{#2}%
    \stepcounter{subsubfigure}%
}{%
    \addtocounter{subfigure}{-1}%
  \end{subfigure}%
}

\newcounter{subsubfigure}

\newlength{\tablesepskip}
\newlength{\tablesubheaderskip}

\fancypagestyle{firstpage}{
  \setlength{\headheight}{3.5\normalbaselineskip}
  \setlength{\headsep}{1.5\normalbaselineskip}

  \fancyhead[C]{%
    \footnotesize%
    \textit{Proc. IEEE MMSP 2023}, Poitiers, France, Sept. 2023 \\
    \vspace{0.8\normalbaselineskip}%
    \scriptsize%
    \copyright{} 2023 IEEE. Personal use of this material is permitted. Permission from IEEE must be obtained for all other uses, in any current or future media, including reprinting/republishing this material for advertising or promotional purposes, creating new collective works, for resale or redistribution to servers or lists, or reuse of any copyrighted component of this work in other works.%
  }
  \fancyfoot{}
}

\begin{document}

\maketitle

\thispagestyle{firstpage}

\begin{abstract}
  Deep learning is increasingly being used to perform machine vision tasks such as classification, object detection, and segmentation on 3D point cloud data.
  However, deep learning inference is computationally expensive.
  The limited computational capabilities of end devices thus necessitate a codec for transmitting point cloud data over the network for server-side processing.
  Such a codec must be lightweight and capable of achieving high compression ratios without sacrificing accuracy.
  Motivated by this, we present a novel point cloud codec that is highly specialized for the machine task of classification.
  Our codec, based on PointNet, achieves a significantly better rate-accuracy trade-off in comparison to alternative methods.
  In particular, it achieves a 94\% reduction in BD-bitrate over non-specialized codecs on the ModelNet40 dataset.
  For low-resource end devices, we also propose two lightweight configurations of our encoder that achieve similar BD-bitrate reductions of 93\% and 92\% with 3\% and 5\% drops in top-1 accuracy, while consuming only 0.470 and 0.048 encoder-side kMACs/point, respectively.
  Our codec demonstrates the potential of specialized codecs for machine analysis of point clouds, and provides a basis for extension to more complex tasks and datasets in the future.
  %
\end{abstract}


\section{Introduction}
\label{sec:introduction}

Point clouds are used to represent 3D visual data in many applications, including autonomous driving, robotics, and augmented reality.
Recent advances in deep learning have led to the development of deep learning-based methods for machine vision tasks on point cloud data.
Common tasks include classification, segmentation, object detection, and object tracking.
However, current deep learning-based methods often require significant computational resources, which impose hardware requirements.
Such significant requirements may not be physically or economically feasible for end devices.

One approach to address the issue of insufficient end-device computational resources is to transmit the point cloud and other sensor data to a server for processing.
However, this introduces its own challenges, including the effects of network availability, latency, and bandwidth.
In order to reduce network requirements, the end device may compress the point cloud data before transmission.
However, network capabilities vary depending on various factors, including end-device location and network congestion.
This means that sufficient bandwidth may still not be available to transmit the point cloud data.
A hybrid strategy is to perform part of the machine task on the end device itself.
This can reduce the amount of data that needs to be transmitted to the server for further processing, without exceeding the computational budget of the end device~\cite{kang2017neurosurgeon}.
This enhances the robustness of the end device to varying network conditions, while potentially improving overall system latency and adaptability~\cite{shlezinger2022IOTM}.

We propose a novel learned point cloud codec for classification.
Our learned codec takes a point cloud as input, and outputs a highly compressed representation that is intended solely for machine analysis.
To our knowledge, this is the first point cloud codec specialized for machine analysis.
Existing codecs for point clouds are designed to reconstruct point clouds intended for human viewing.
This means that a significant amount of bits is wasted on encoding information that is not strictly relevant to machine analysis.
By partially processing the point cloud before compression, our codec is able to achieve significantly better compression performance, without compromising task accuracy.

In this paper, we present our task-specialized codec architecture in full and lightweight configurations.
We evaluate its rate-accuracy (RA) performance on the ModelNet40 dataset~\cite{wu20143d}.
We also investigate how the number of input points (and thus reduced computation) affects the RA performance of our proposed codec.
Furthermore, we compare our proposed codec's performance against alternative non-specialized methods.
Our code for training and evaluation is available online%
\footnote{%
  \hfill%
  \url{https://github.com/multimedialabsfu/learned-point-cloud-compression-for-classification}%
}%
.

\section{Related work}
\label{sec:related-work}

Point cloud classification models can be organized into groups based on the type of input data they accept.
Models such as VoxNet~\cite{maturana2015voxnet} take as input point clouds that have been preprocessed into a voxel grid.
Unfortunately, these methods often use 3D convolutions, which require a significant amount of computational resources.
Additionally, since most voxels are usually empty, these methods arguably waste a significant amount of computation on empty space.
Furthermore, the voxel grid representation is not very compact, and thus requires a significant amount of memory for higher spatial resolutions (e.g., a 32-bit tensor of shape $1024 \times 1024 \times 1024$ occupies 32 GB).
Models such as OctNet~\cite{riegler2016octnet} take octrees as input.
Octrees offer a more compact representation of the voxelized point cloud by encoding the node occupancy in bitstrings.
Large unoccupied regions of space may be represented via a single "0" node in an octree.
Point-based models such as PointNet~\cite{qi2016pointnet} and PointNet++~\cite{qi2017pointnetplusplus} directly accept raw point lists $(x_1, x_2, \ldots, x_P)$, where $x_i \in \mathbb{R}^3$ represents a point in a 3D space and $P$ is the number of points.
Some challenges faced with this input format include designing order-invariant models (due to the lack of a worthwhile canonical ordering of points),
as well as in devising operations capable of using the metric structure induced by point locality.
Despite the challenges, point-based models are able to achieve surprisingly competitive accuracy,
and offer the most promise in terms of minimizing computational requirements.


PointNet~\cite{qi2016pointnet}, which our proposed architecture is based on, can be represented as an input permutation-invariant function:
\[ f(x_1, \ldots, x_n) = (\gamma \circ \pi)(h(x_1), \ldots, h(x_n)), \]
where $h$ is applied to each point $x_i$ individually, $\pi$ is a simple permutation-invariant function, and $\gamma$ may be any function.
In the original PointNet architecture, $h$ is a weight-shared MLP, $\pi$~is a max pooling function applied across the point dimension, and $\gamma$ is an MLP.

In the related field of learned image compression, Ball{\'e} \emph{et al.}~\cite{balle2018variational} proposed a variational autoencoder (VAE) architecture for image compression.
Here, the model transforms the input $\boldvar{x}$ into a latent representation $\boldvar{y}$, which is then quantized into $\boldvar{\hat{y}}$ and losslessly compressed using a learned entropy model.
The codec is trained end-to-end using the loss function
\begin{equation}
    \mathcal{L} = R + \lambda \cdot D(\boldvar{x}, \boldvar{\hat{x}}),
    \label{eq:rd-loss}
\end{equation}
where $D(\boldvar{x}, \boldvar{\hat{x}})$ is the distortion measure between input $\boldvar{x}$ and decoded $\boldvar{\hat{x}}$, and $R$ is the estimate of the entropy of $\boldvar{\hat{y}}$.
One simple entropy model, known in the literature as an \emph{entropy bottleneck}, makes a "factorized" prior assumption --- that each element within a latent channel is independently and identically distributed.
It models a monotonically increasing non-parametric cumulative distribution function using a differentiable MLP.
This mechanism has also shown effectiveness in learned codecs in other fields, including learned point cloud compression (PCC), and has been incorporated by a variety of works including~\cite{yan2019deep,he2022density,pang2022graspnet,fu2022octattention,you2022ipdae}.

We have also based our work on ideas introduced for machine tasks on images.
Early works demonstrated the use of standard codecs in compressing the latent features~\cite{choi2018mmsp}.
More recently, approaches such as Video Coding for Machines (VCM)~\cite{duan2020vcm} and Coding for Machines (CfM) have gained traction in the research community.
For instance, works such as~\cite{hu2020towardscfhmvscalable,choi2022sichm} demonstrate the potential bitrate savings of scalable image compression in a multi-task scenario for a machine vision task (e.g., facial landmark detection, or object detection) and human vision (e.g., image reconstruction).
In this work, we focus solely on a single machine vision task applied to point cloud data.

\section{Proposed codec}
\label{sec:proposed-codec}

\subsection{Input compression}

\begin{figure}[tbp]
  \centering
  \begin{subfigure}[b]{0.68\linewidth}
    \centering
    \includegraphics[width=\linewidth]{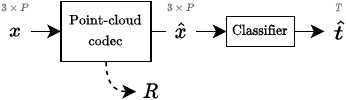}
    \caption{input compression}
    \label{fig:arch-comparison/input-compression}
  \end{subfigure}%
  \vspace{1.5\baselineskip}
  \begin{subfigure}[b]{0.8\linewidth}
    \centering
    \includegraphics[width=\linewidth]{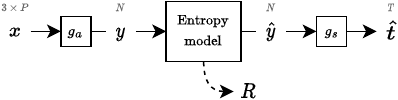}
    \caption{proposed}
    \label{fig:arch-comparison/proposed}
  \end{subfigure}%
  \caption{High-level comparison of codec architectures.}
  \label{fig:arch-comparison}
\end{figure}

In \cref{fig:arch-comparison/input-compression}, we show an abstract representation of an input codec, similar to the "chain" configuration explored by~\cite{chamain2020endtoend} for end-to-end image compression for machines.
In this codec, the input point cloud $\boldvar{x}$ is encoded directly, without any intermediate feature extraction.
On the decoder side, the point cloud is then reconstructed as $\boldvar{\hat{x}}$.
Any point cloud compression codec can be used for this purpose, including standard non-learned codecs such as G-PCC~\cite{mpeg2019gpccv2}.
Finally, the reconstructed point cloud $\boldvar{\hat{x}}$ is fed into a classification model (e.g., PointNet) in order to obtain the class prediction $\boldvar{\hat{t}}$.
This approach provides a baseline for comparison with our proposed codec.

\subsection{Motivation for the proposed codec}

An efficient task-specific codec can be developed using the concept of Information Bottleneck (IB)~\cite{IB_Allerton1999}:
\begin{equation}
  \min_{p(\boldvar{\hat{y}} \mid \boldvar{x})} \quad I(\boldvar{x}; \boldvar{\hat{y}}) - \beta \cdot I(\boldvar{\hat{y}} ; \boldvar{\hat{t}}),
\label{eq:IB}
\end{equation}
where $I(\cdot;\cdot)$ is the mutual information~\cite{Cover_Thomas_2006}, $p(\boldvar{\hat{y}} \mid \boldvar{x})$ is the mapping from the input point cloud $\boldvar{x}$ to the latent representation $\boldvar{\hat{y}}$, and $\beta>0$ is the IB Lagrange multiplier~\cite{IB_Allerton1999}.
We can think of $p(\boldvar{\hat{y}} \mid \boldvar{x})$ as feature extraction followed by quantization.
Hence, $\boldvar{\hat{y}}$ is fully determined whenever $\boldvar{x}$ is given, so $H(\boldvar{\hat{y}} \mid \boldvar{x}) = 0$, where $H(\cdot \mid \cdot)$ is the conditional entropy~\cite{Cover_Thomas_2006}.
Therefore, $I(\boldvar{x}; \boldvar{\hat{y}}) = H(\boldvar{\hat{y}}) - H(\boldvar{\hat{y}} \mid \boldvar{x}) = H(\boldvar{\hat{y}})$.

\begin{figure*}[tbp]
  \centering
  \includegraphics[width=0.7\linewidth]{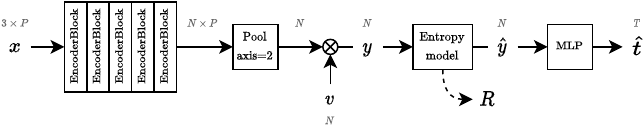}
  \caption{Proposed codec architecture.}
  \label{fig:arch-proposed-full}
\end{figure*}

Furthermore, since decreasing $-\beta\cdot I(\boldvar{\hat{y}} ; \boldvar{\hat{t}})$ would improve the accuracy of the task, we can use $\lambda \cdot D(\boldvar{t},\boldvar{\hat{t}})$ as a proxy for $-\beta\cdot I(\boldvar{\hat{y}} ; \boldvar{\hat{t}})$, where $\boldvar{t}$ is the ground-truth label, $\boldvar{\hat{t}}$ is the label produced using the compressed latent representation, and $D(\boldvar{t},\boldvar{\hat{t}})$ is a distortion measure.
Therefore, in our case, the IB~\eqref{eq:IB} becomes:
\begin{equation}
  \min_{p(\boldvar{\hat{y}} \mid \boldvar{x})} \quad H(\boldvar{\hat{y}}) + \lambda \cdot D(\boldvar{t},\boldvar{\hat{t}}).
\label{eq:IB_task}
\end{equation}
It is clear that the form of IB in~\eqref{eq:IB_task} is analogous to the loss function~\eqref{eq:rd-loss}.
We make use of this analogy to develop the proposed codec, which is described next.

\subsection{Proposed architecture}

In \cref{fig:arch-comparison/proposed}, we show a high-level representation of our proposed codec architecture.
Following the terminology of~\cite{balle2018variational}, we refer to $g_a$ as the \emph{analysis transform}, and $g_s$ as the \emph{synthesis transform}.
In this architecture, the input point cloud $\boldvar{x}$ is first encoded into a latent representation $\boldvar{y} = g_a(\boldvar{x})$, which is then quantized as $\boldvar{\hat{y}} = Q(\boldvar{y})$, and then losslessly compressed using a learned entropy model.
Therefore, $p(\boldvar{\hat{y}} \mid \boldvar{x})$ from the IB is $Q \circ g_a$.
Then, the reconstructed latent representation $\boldvar{\hat{y}}$ is used to predict the classes $\boldvar{\hat{t}} = g_s(\boldvar{\hat{y}})$.

Our proposed architecture, visualized in~\cref{fig:arch-proposed-full}, is based on the PointNet~\cite{qi2016pointnet} classification model.
The input 3D point cloud containing $P$ points is represented as a matrix $\boldvar{x} \in \mathbb{R}^{3 \times P}$.
The input $\boldvar{x}$ is fed into a sequence of encoder blocks.
Each encoder block consists of a convolutional layer, a batch normalization layer, and a ReLU activation.
As described in~\cite{qi2016pointnet}, the early encoder blocks must be applied to each input point independently and identically, in order to avoid learning input permutation-specific dependencies.
Therefore, we use pointwise convolutional layers with kernel size 1, which are exactly equivalent to the "shared MLP" described in~\cite{qi2016pointnet}.

Following the sequence of encoder blocks, a max pooling operation is applied along the point dimension to generate an input permutation-invariant feature vector of length $N$.
The resulting feature vector is then multiplied element-wise by a trainable gain vector $v \in \mathbb{R}^{N}$, which is initialized to $[1, 1, \ldots, 1]$ before training.
Due to the batch normalization layers, the resulting feature vector has small values.
Thus, in order to improve training stability and the rate of convergence, the feature vector is multiplied by a constant scalar value of $10$.
The resulting vector is the output of the encoder-side analysis transform, which we label as $\boldvar{y}$.


Then, we quantize $\boldvar{y}$ via uniform quantization (specifically, integer rounding) to obtain $\boldvar{\hat{y}}$.
During training, uniform quantization is simulated using additive uniform noise $\mathcal{U}(-0.5, 0.5)$.
The quantized vector $\boldvar{\hat{y}}$ is then losslessly encoded using a fully-factorized learned entropy model introduced in~\cite{balle2018variational}.

On the decoder side, the decoded vector $\boldvar{\hat{y}}$ is fed into an MLP consisting of fully-connected layers interleaved with ReLU activations and batch normalizations.
Before the last fully-connected layer, we use a dropout layer that randomly sets 30\% of its inputs to zero.
The output of the MLP is a vector of logits $\boldvar{\hat{t}} \in \mathbb{R}^T$, where $T$ is the number of classes.

We provide "full", "lite", and "micro" configurations of the proposed codec.
For each configuration, \cref{tbl:layers} lists the number of layer output channel sizes along with the estimated MAC (multiply-accumulate) counts%
\footnote{One MAC operation may be considered equivalent to a FLOP (floating-point operation) on most hardware.}%
.
Group convolutional layers are specified in the format "output\_size/group".
In contrast to~\cite{qi2016pointnet}, we do not use any input or feature transformations in order to simplify the architectures for clearer analysis, as well as to reduce the computational requirements.

%
%

\begin{table}[t]
  \centering
  \caption{Layer sizes and MAC counts for various proposed codec types}
  \label{tbl:layers}
  %
  \scriptsize
  %
  \setlength{\tablesepskip}{-0.9\normalbaselineskip}
  \begin{tabular}[]{ccccc}
    \toprule
    \\[\tablesepskip]
    Proposed           & Encoder           & Decoder                      & Encoder            & Decoder   \\
    codec              & layer sizes       & layer sizes                  & MAC/pt             & MAC       \\
    \\[\tablesepskip]
    \midrule
    \\[\tablesepskip]
    full               & 64 64 64 128 1024 & 512 256 40                   & 150k               & 670k      \\
    lite               & 8 8 16 16/2 32/4  & 512 256 40                   & 0.47k              & 160k      \\
    micro              & 16                & 512 256 40                   & 0.048k             & 150k      \\
    \\[\tablesepskip]
    \bottomrule
  \end{tabular}
\end{table}

\subsection{Lightweight and micro architectures}

In addition to our "full" proposed codec, we also provide a lightweight configuration, which we denote as "lite".
In this architecture, the encoder-side layers contain fewer output channels.
To further reduce encoder-side computational costs, they also use group convolutions with channel shuffles in between, as is done in ShuffleNet~\cite{zhang2017shufflenet}.
After training, the gain and batch normalization layers may be fused into the preceding convolutional layer.
The "lite" architecture strikes a balance between RA performance and encoder complexity.
In fact, the encoder-side transform requires only 0.47k MACs/point, which is significantly less than the 150k MACs/point required by the "full" architecture encoder.
For input point clouds consisting of $P=256$ points, the total MAC count for the "lite" codec is 120k, which is below the corresponding decoder-side MAC count of 160k.  

Additionally, we examine a "micro" architecture, whose encoder-side transform consists of only a single encoder block with 16 output channels, and a max pooling operation.
This codec is useful for analysis and comparison --- and yet, it is also capable of surprisingly competitive RA performance.

\section{Experiments}
\label{sec:experiments}

Our models were trained on the ModelNet40~\cite{wu20143d} dataset, which consists of 12311 different 3D object models organized into 40 classes.
We used an Adam optimizer with a learning rate of 0.001.
Our code was written using the PyTorch, CompressAI~\cite{begaint2020compressai}, and CompressAI Trainer~\cite{ulhaq2022compressaitrainer} libraries.

The loss function that is minimized during training is:
\[
  \mathcal{L} = R + \lambda \cdot D(\boldvar{t}, \boldvar{\hat{t}}),
\]
where the rate $R = -\log p_{\boldvar{\hat{y}}}(\boldvar{\hat{y}})$ is the log of the likelihoods outputted by the entropy model, and
the distortion $D(\boldvar{t}, \boldvar{\hat{t}})$ is the cross-entropy between the one-hot encoded labels $\boldvar{t}$ and the softmax of the model's prediction $\boldvar{\hat{t}}$.
We trained different models to operate at different rate points by varying the hyperparameter $\lambda \in [10, 16000]$.

\subsection{Proposed codec}

For each of the proposed codec architectures, we trained a variation of each codec to accept an input point cloud containing $P \in \mathcal{P}$ points.
We trained eight such variations for each of the values in the set
$\mathcal{P} = \{ 8, 16, 32, 64, 128, 256, 512, 1024 \}$.
Although each codec is capable of handling a variable number of points,
training a separate model for each $P$ guarantees that each codec is well-optimized for each rate-accuracy trade-off.

\subsection{Input compression codec}

We compare our proposed codec against an "input compression" codec architecture.
For this codec, the encoder may be taken from any point cloud codec.
We have tested multiple codecs, including TMC13~\cite{mpeg2021tmc13} v14 (an implementation of the G-PCC v2~\cite{mpeg2019gpccv2} standard), OctAttention~\cite{fu2022octattention}, and Draco~\cite{google2017draco}.
On the decoder side is the corresponding point cloud decoder, followed by a PointNet classification model.
We trained a PointNet model (without the input and feature transforms) for each $P \in \mathcal{P}$.

We generated eight separate datasets of $P$-point point clouds, where each point cloud was uniformly subsampled from the test dataset.
Then, we compressed and decompressed each point cloud from each $P$-point dataset at various compression ratios.
The compression ratio can be effectively controlled by varying the amount of input scaling, which we denote by $S$.
(The input scaling parameter is directly proportional to the number of bins used during uniform quantization of the input points.)
We varied $S$ over the set $\mathcal{S} = \{1, 2, 4, \ldots, 256\}$ and $P$ over $\mathcal{P}$ to produce $|\mathcal{P}| \cdot |\mathcal{S}|$ distinct datasets.
We evaluated each dataset associated with the pair $(P, S) \in \mathcal{P} \times \mathcal{S}$ on the correspondingly trained PointNet models to obtain a set of rate-accuracy points.
Finally, we took the Pareto front of this set to obtain the best rate-accuracy curve achieved by the tested input compression codec.

\subsection{Reconstruction}

In order to visually assess the contents of the machine task-specialized bitstream, we trained a point cloud reconstruction network on top of our trained models.
This auxiliary network was trained to minimize the loss function
$\mathcal{L} = D(\boldvar{x}, \boldvar{\hat{x}})$,
where we used Chamfer distance for $D$.

We also identify a critical point set for a fixed point cloud.
A critical point set is a minimal set of points which generate the exact same latent $\boldvar{y}$, and correspondingly, the same bitstream.
Formally, for any given point cloud $\boldvar{x}$, let $\boldvar{x}_C \subseteq \boldvar{x}$ denote a (not necessarily unique) critical point set.
Then, $g_a(\boldvar{x}_C) = g_a(\boldvar{x}) = \boldvar{y}$, and there is uniquely one valid critical point set $(\boldvar{x}_C)_C$ for $\boldvar{x}_C$, and it is itself.
Since $g_a$ contains a max pooling operation, the critical point set is not theoretically unique; however, in practice, it is rare for there to be more than one critical point set.
A critical point set may be computed by
$\boldvar{x}_C = \bigcup_{1 \leq j \leq N} \argmax_{\boldvar{x}_i \in \boldvar{x}} \, (h(\boldvar{x}_i))_j$,
where $\{h(\boldvar{x}_i) : 1 \leq i \leq P\}$ represents the entire set of generated latent vectors immediately preceding max pooling.

\section{Results}
\label{sec:results}

\begin{figure*}[htbp]
  \centering
  \newcommand{\subfigurehspace}{.25\linewidth}
  \begin{subfigure}[b]{\subfigurehspace}
    \includegraphics[width=\linewidth]{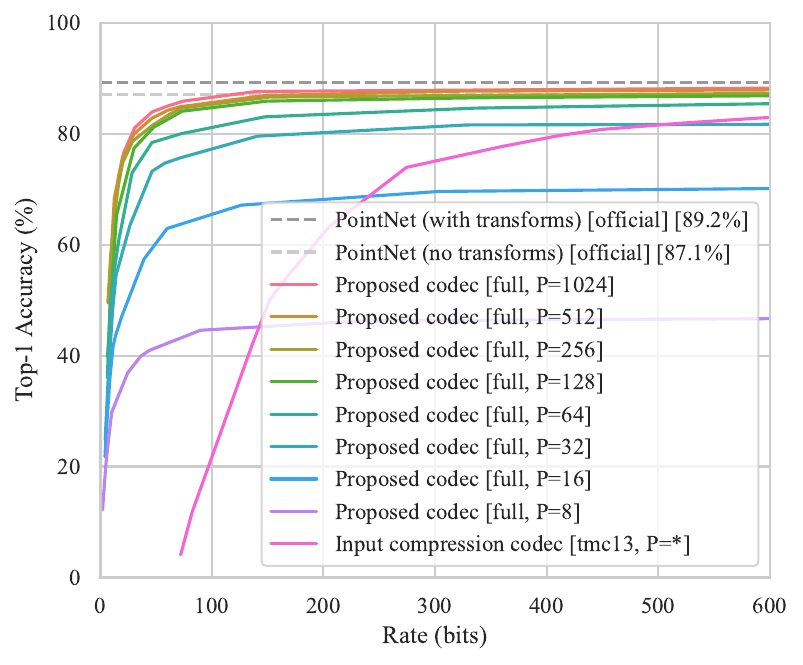}
    \caption{``full'' codec}
    \label{fig:rate-accuracy/full}
  \end{subfigure}%
  \hfill%
  \begin{subfigure}[b]{\subfigurehspace}
    \centering
    \includegraphics[width=\linewidth]{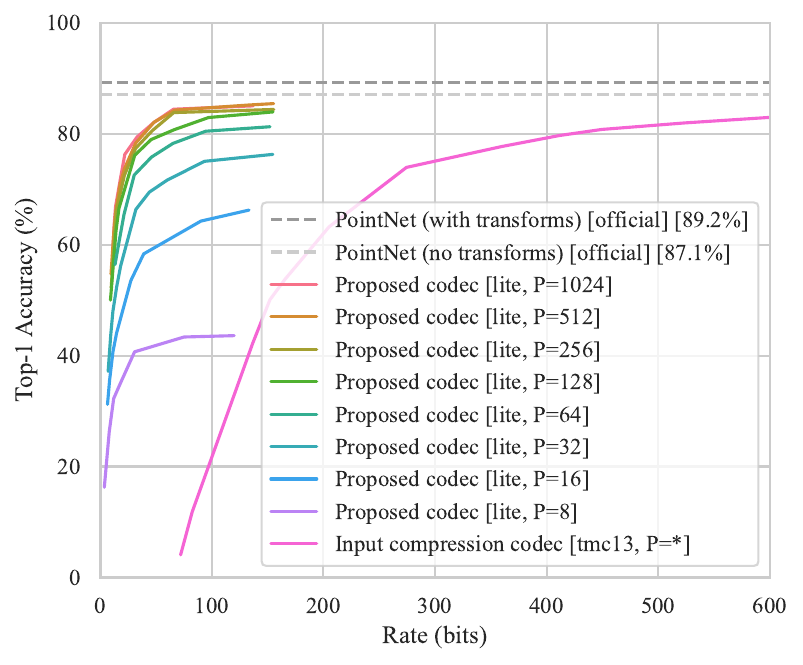}
    \caption{``lite'' codec}
    \label{fig:rate-accuracy/lite}
  \end{subfigure}%
  \hfill%
  \begin{subfigure}[b]{\subfigurehspace}
    \centering
    \includegraphics[width=\linewidth]{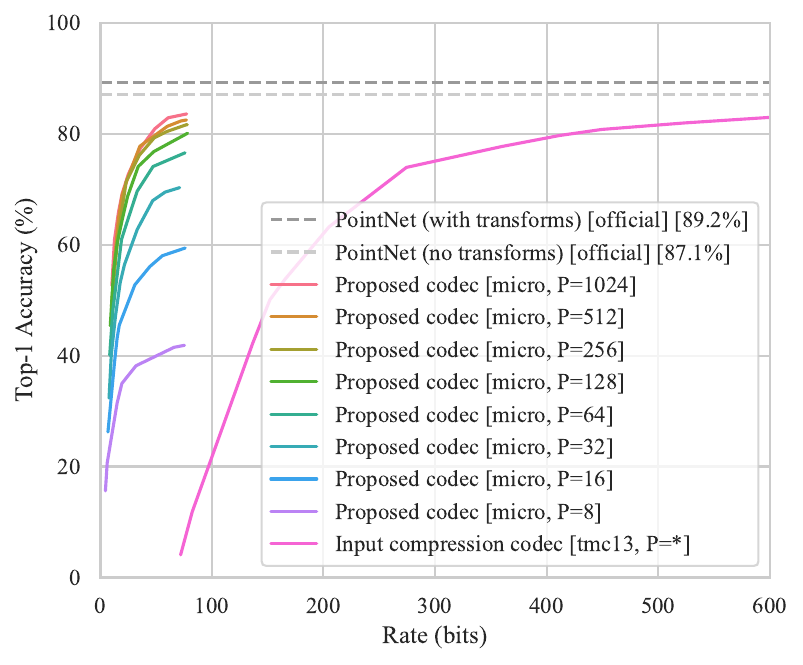}
    \caption{``micro'' codec}
    \label{fig:rate-accuracy/micro}
  \end{subfigure}%
  \hfill%
  \begin{subfigure}[b]{\subfigurehspace}
    \centering
    \includegraphics[width=\linewidth]{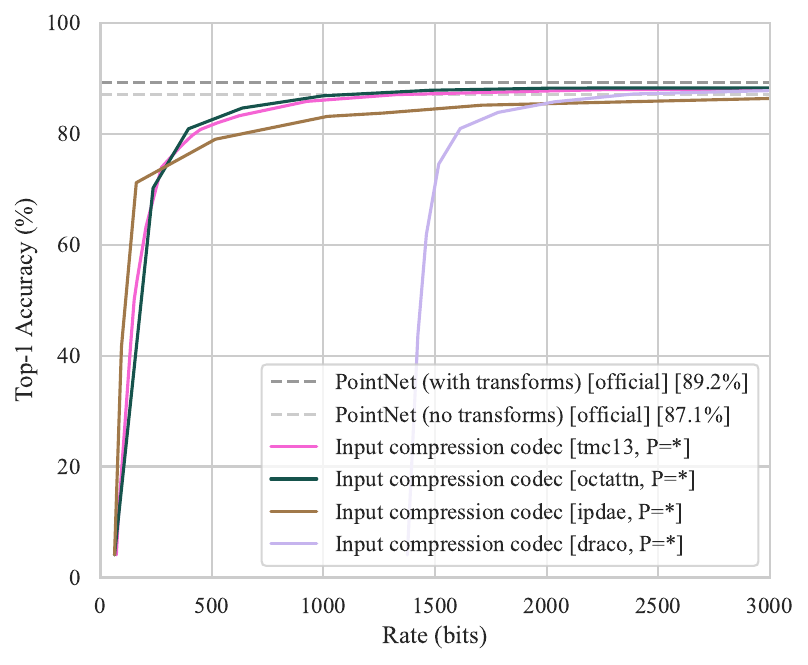}
    \caption{input compression codecs}
    \label{fig:rate-accuracy/input-compression}
  \end{subfigure}%
  \caption{Rate-accuracy curves evaluated on the ModelNet40 test set.}
  \label{fig:rate-accuracy}
\end{figure*}

\cref{fig:rate-accuracy} shows the rate-accuracy (RA) curves for the proposed "full", "lite", and "micro" codecs in comparison with the input compression codec.
Also included are two baseline accuracies taken from the official PointNet paper~\cite{qi2016pointnet}, for the model with (89.2\%) and without (87.1\%) the input/feature transforms.
Since our compression models were all trained \emph{without} the input/feature transforms, the lower baseline offers a more direct comparison.
In \cref{tbl:measurements}, we list the peak accuracies attained by each codec, as well as the Bjøntegaard-Delta (BD)~\cite{bjontegaard2001calculation} improvements in rates and accuracies relative to the reference input compression codec.

\begin{table}[t]
  \centering
  \begin{threeparttable}
  \caption{BD metrics and max attainable accuracies per codec}
  \label{tbl:measurements}
  \scriptsize
  %
  \newcommand{\tablesubheaderstyle}[1]{{\underline{#1}}}
  \newcommand{\tbleq}[1]{{\normalsize \scalebox{0.7}{{#1}}}}
  \setlength{\tablesepskip}{-0.8\normalbaselineskip}
  \setlength{\tablesubheaderskip}{0.6\normalbaselineskip}
  \begin{tabular}[]{lccc}
    \toprule
    \\[\tablesepskip]
    Codec                   & Max acc (\%) & BD rate (rel \%) & BD acc (\%)
    \\
    \\[\tablesepskip]
    \midrule
    \\[\tablesepskip]
    \tablesubheaderstyle{Input compression} \\[\tablesubheaderskip]
    TMC13~\cite{mpeg2021tmc13}                  &   88.5 &    0.0 &    0.0 \\
    OctAttention~\cite{fu2022octattention}      &   88.4 &  -13.2 &   +2.1 \\
    IPDAE~\cite{you2022ipdae}                   &   87.0 &  -23.0 &   +3.6 \\
    Draco~\cite{google2017draco}                &   88.3 & +780.7 &   -4.2 \\
    \\[\tablesepskip]
    \midrule
    \\[\tablesepskip]
    \tablesubheaderstyle{Proposed (full)} \\[\tablesubheaderskip]
    \tbleq{$P=1024$}        &   88.5 &  -93.8 &  +16.4 \\
    \tbleq{$P=512$}         &   88.0 &  -93.7 &  +15.9 \\
    \tbleq{$P=256$}         &   87.6 &  -93.3 &  +15.4 \\
    \tbleq{$P=128$}         &   87.1 &  -92.7 &  +14.9 \\
    \tbleq{$P=64$}          &   86.1 &  -91.1 &  +13.2 \\
    \tbleq{$P=32$}          &   81.8 &  -90.6 &   +9.3 \\
    \tbleq{$P=16$}          &   70.4 &  -86.8 &   -2.3 \\
    \tbleq{$P=8$}           &   46.8 &  -88.5 &  -25.3 \\
    \\[\tablesepskip]
    \midrule
    \\[\tablesepskip]
    \tablesubheaderstyle{Proposed (lite)} \\[\tablesubheaderskip]
    \tbleq{$P=1024$}        &   85.0 &  -93.0 &  +13.5 \\
    \tbleq{$P=512$}         &   85.5 &  -92.8 &  +14.2 \\
    \tbleq{$P=256$}         &   84.4 &  -92.4 &  +12.8 \\
    \tbleq{$P=128$}         &   84.0 &  -91.6 &  +12.5 \\
    \tbleq{$P=64$}          &   81.3 &  -88.5 &   +9.8 \\
    \tbleq{$P=32$}          &   76.3 &  -88.7 &   +4.9 \\
    \tbleq{$P=16$}          &   66.2 &  -86.1 &   -4.1 \\
    \tbleq{$P=8$}           &   43.6 &  -90.2 &  -28.0 \\
    \\[\tablesepskip]
    \midrule
    \\[\tablesepskip]
    \tablesubheaderstyle{Proposed (micro)} \\[\tablesubheaderskip]
    \tbleq{$P=1024$}        &   83.6 &  -91.8 &  +12.7 \\
    \tbleq{$P=512$}         &   82.5 &  -91.6 &  +11.6 \\
    \tbleq{$P=256$}         &   81.6 &  -91.1 &  +11.0 \\
    \tbleq{$P=128$}         &   80.1 &  -90.9 &   +9.9 \\
    \tbleq{$P=64$}          &   76.6 &  -89.9 &   +6.5 \\
    \tbleq{$P=32$}          &   70.3 &  -89.0 &   +0.1 \\
    \tbleq{$P=16$}          &   59.4 &  -87.6 &  -10.8 \\
    \tbleq{$P=8$}           &   41.9 &  -88.3 &  -28.8 \\
    \\[\tablesepskip]
    \bottomrule
  \end{tabular}
  \begin{tablenotes}
    \item $P$ is the number of points in the input $\boldvar{x}$.
      The BD metrics were computed using the TMC13 input compression codec as the reference anchor.
  \end{tablenotes}
  \end{threeparttable}
\end{table}


Our "full" proposed codec, which is an extension of PointNet (no transforms), achieves the lower baseline accuracy at $120$ bits, and an 80\% accuracy at $30$ bits.
Our "lite" proposed codec saturates in RA performance at around $P=512$ input points.
At around $P=256$, the total MAC count of the proposed "lite" encoder is roughly equal to the decoder.
As shown by the rate-accuracy curves, the $P=256$ model does not suffer too significant a drop in rate-accuracy performance.
This suggests that our method is capable of achieving a good trade-off between rate, accuracy, and runtime performance.
Similarly, our "micro" codec suffers a further slight drop in rate-accuracy performance, but achieves another significant improvement in runtime performance.
The input compression codec is the worst performing codec, and attains the lower baseline accuracy at roughly 2000 bits.

\begin{figure}[t]
  \centering
  \newcommand{\subfigureouterhspace}{\linewidth}
  \newcommand{\subfigurehspace}{.234\linewidth}
  \begin{subfigure}[b]{\subfigureouterhspace}
    \setcounter{subsubfigure}{0}
    \centering
    \begin{subsubfigure}[b]{\subfigurehspace}
      \centering
      \includegraphics[width=\linewidth]{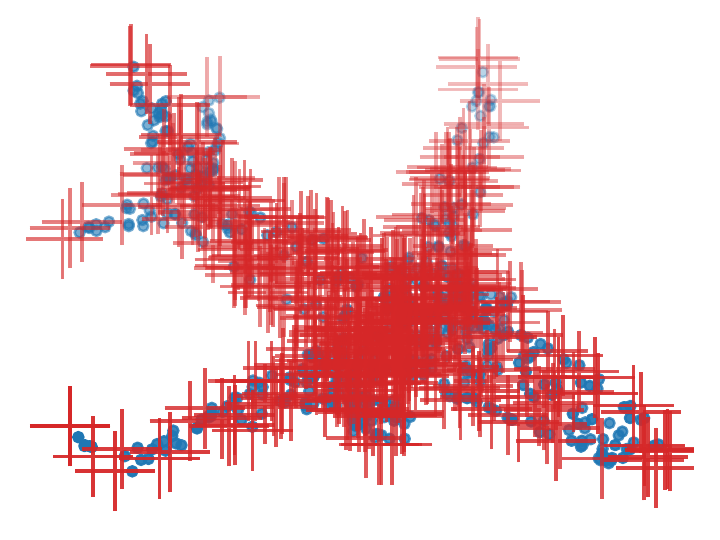}
      \includegraphics[width=\linewidth]{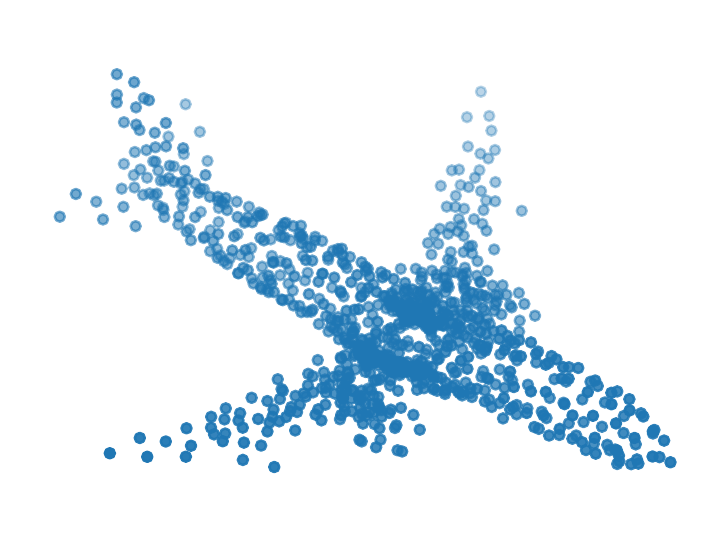}
      \caption{284 bits}
      \label{fig:rec/full/4}
    \end{subsubfigure}%
    \hfill%
    \begin{subsubfigure}[b]{\subfigurehspace}
      \centering
      \includegraphics[width=\linewidth]{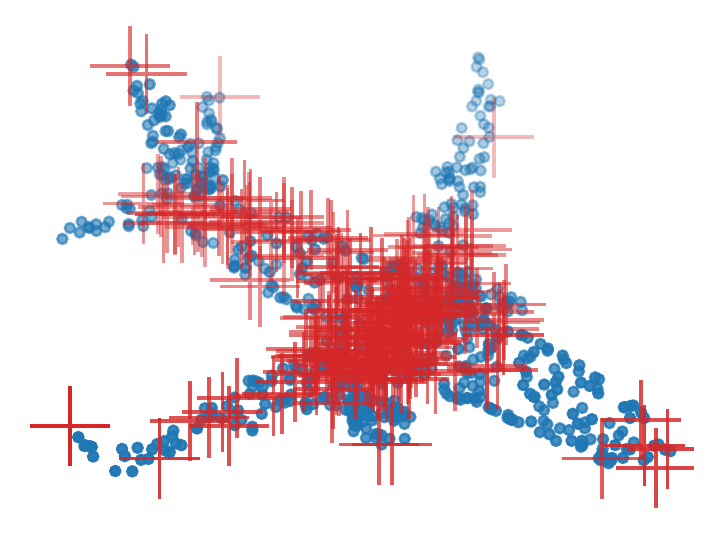}
      \includegraphics[width=\linewidth]{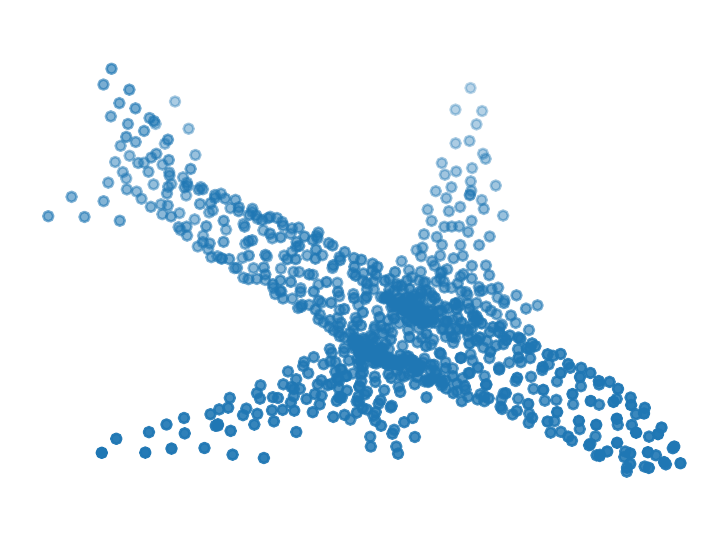}
      \caption{43 bits}
      \label{fig:rec/full/3}
    \end{subsubfigure}%
    \hfill%
    \begin{subsubfigure}[b]{\subfigurehspace}
      \centering
      \includegraphics[width=\linewidth]{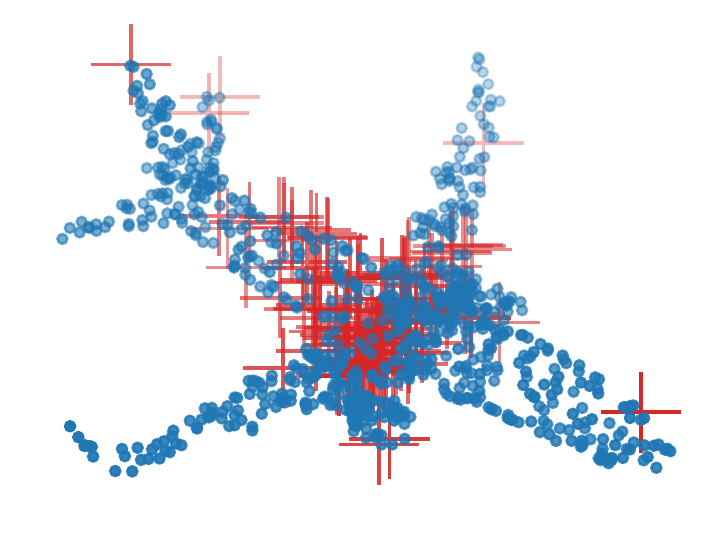}
      \includegraphics[width=\linewidth]{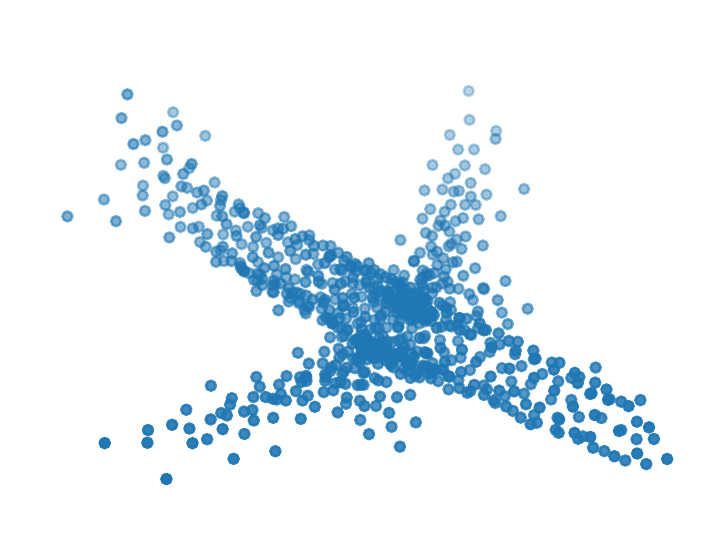}
      \caption{18 bits}
      \label{fig:rec/full/2}
    \end{subsubfigure}%
    \hfill%
    \begin{subsubfigure}[b]{\subfigurehspace}
      \centering
      \includegraphics[width=\linewidth]{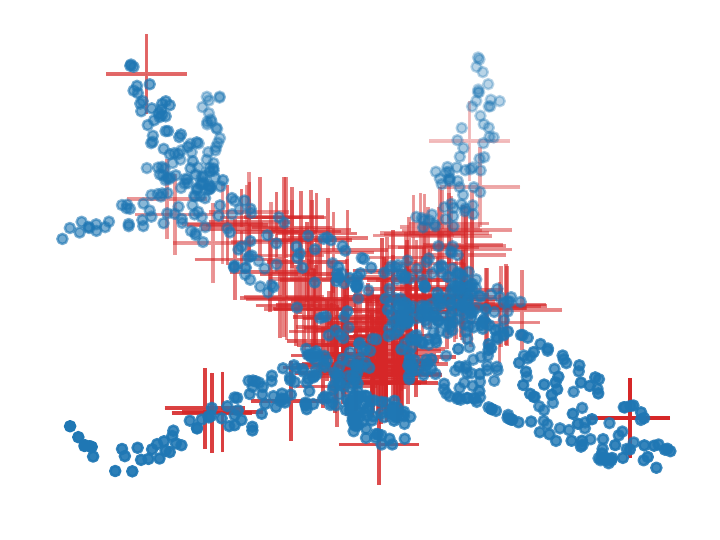}
      \includegraphics[width=\linewidth]{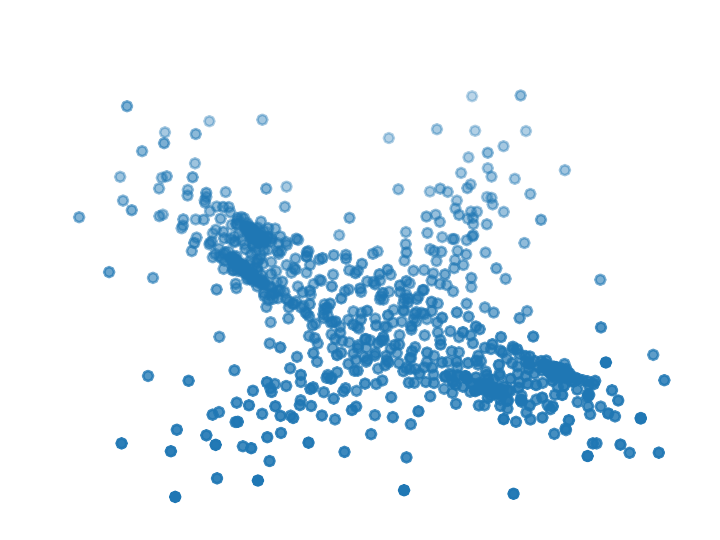}
      \caption{12 bits}
      \label{fig:rec/full/1}
    \end{subsubfigure}%
    \caption{"full" codec}
  \end{subfigure}%
  \par%
  \vspace{1\baselineskip}%
  \begin{subfigure}[b]{\subfigureouterhspace}
    \setcounter{subsubfigure}{0}
    \centering
    \begin{subsubfigure}[b]{\subfigurehspace}
      \centering
      \includegraphics[width=\linewidth]{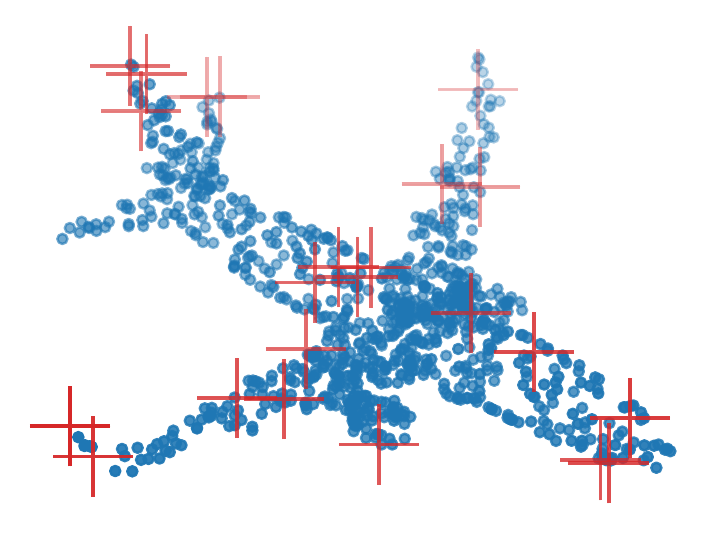}
      \includegraphics[width=\linewidth]{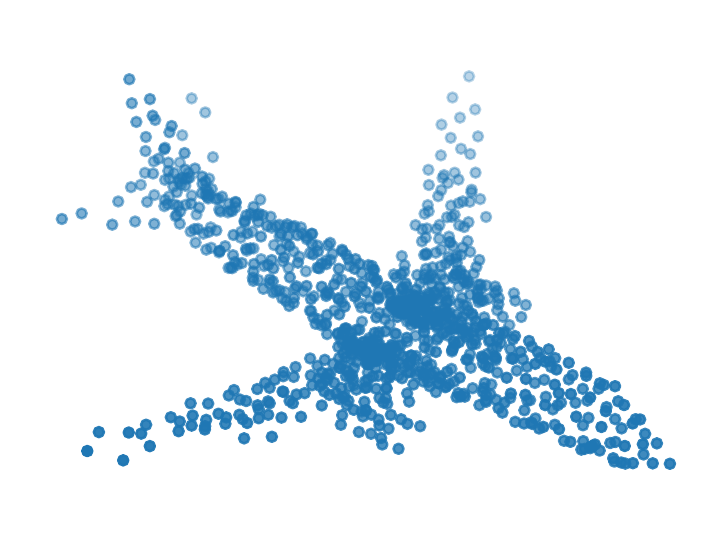}
      \caption{124 bits}
      \label{fig:rec/lite/4}
    \end{subsubfigure}%
    \hfill%
    \begin{subsubfigure}[b]{\subfigurehspace}
      \centering
      \includegraphics[width=\linewidth]{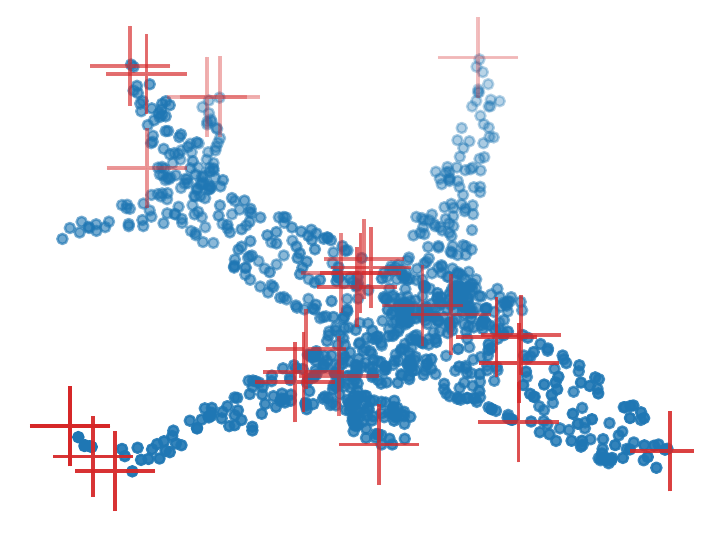}
      \includegraphics[width=\linewidth]{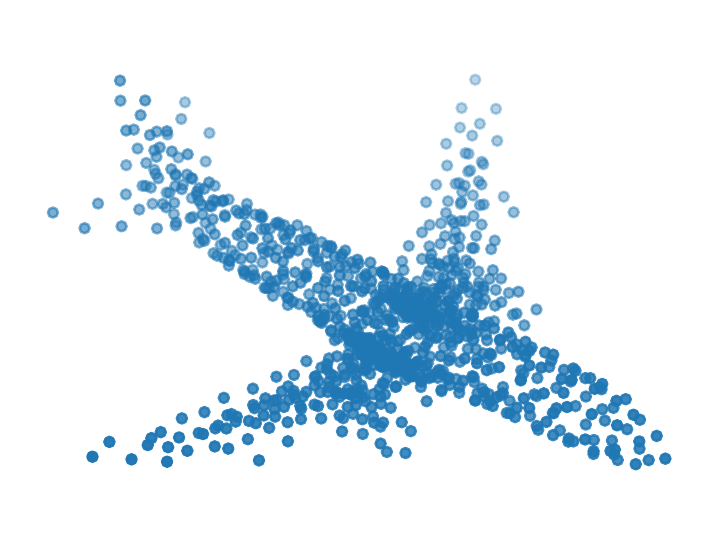}
      \caption{36 bits}
      \label{fig:rec/lite/3}
    \end{subsubfigure}%
    \hfill%
    \begin{subsubfigure}[b]{\subfigurehspace}
      \centering
      \includegraphics[width=\linewidth]{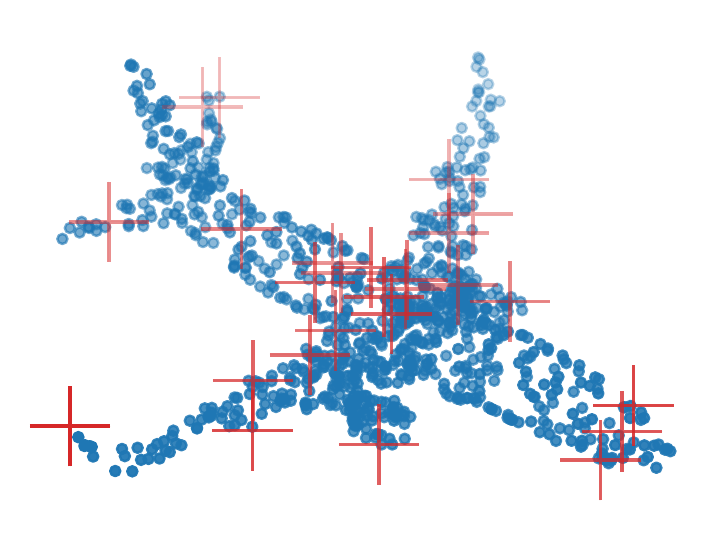}
      \includegraphics[width=\linewidth]{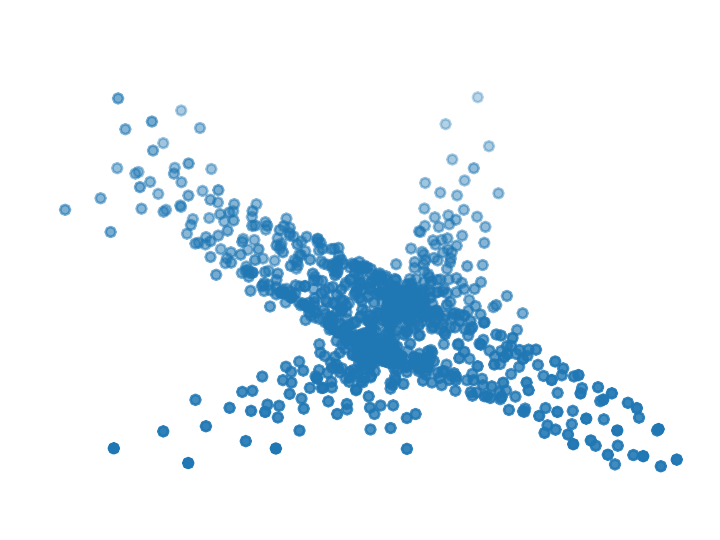}
      \caption{19 bits}
      \label{fig:rec/lite/2}
    \end{subsubfigure}%
    \hfill%
    \begin{subsubfigure}[b]{\subfigurehspace}
      \centering
      \includegraphics[width=\linewidth]{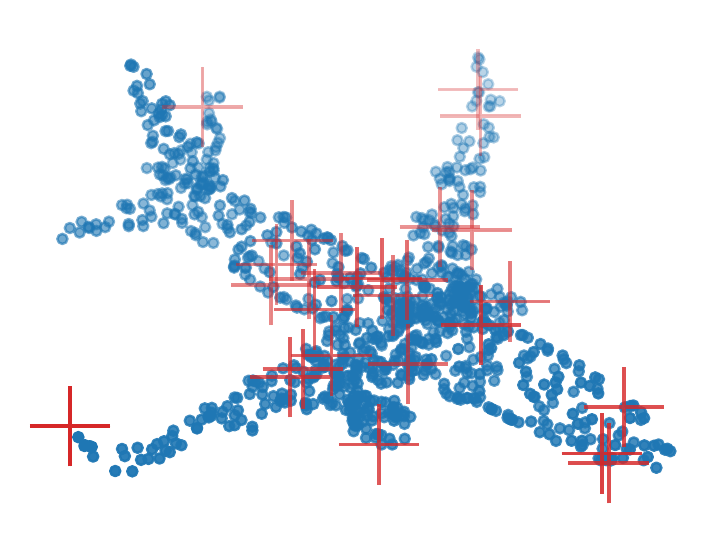}
      \includegraphics[width=\linewidth]{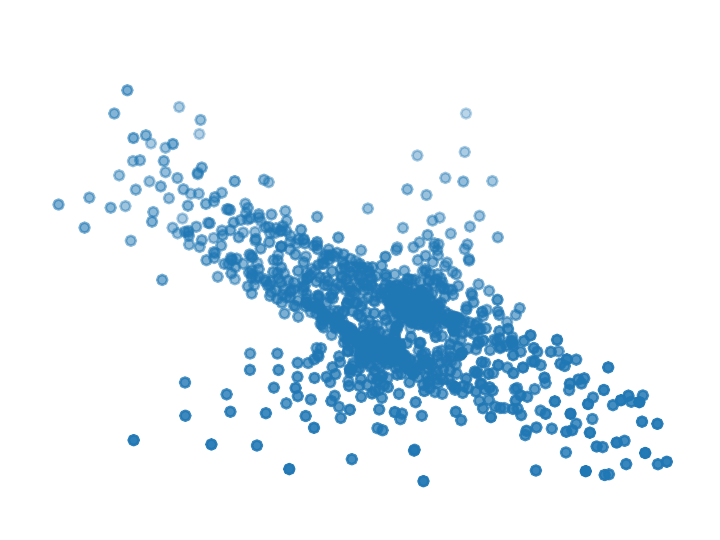}
      \caption{13 bits}
      \label{fig:rec/lite/1}
    \end{subsubfigure}%
    \caption{"lite" codec}
  \end{subfigure}%
  \par%
  \vspace{1\baselineskip}%
  \begin{subfigure}[b]{\subfigureouterhspace}
    \setcounter{subsubfigure}{0}
    \centering
    \begin{subsubfigure}[b]{\subfigurehspace}
      \centering
      \includegraphics[width=\linewidth]{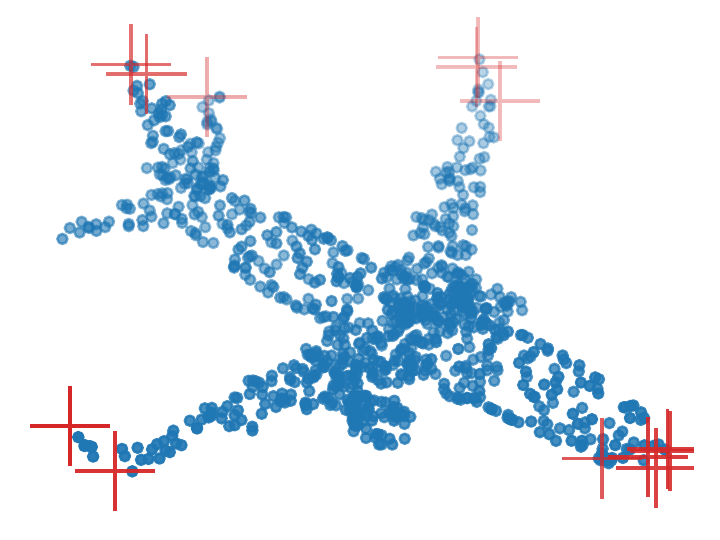}
      \includegraphics[width=\linewidth]{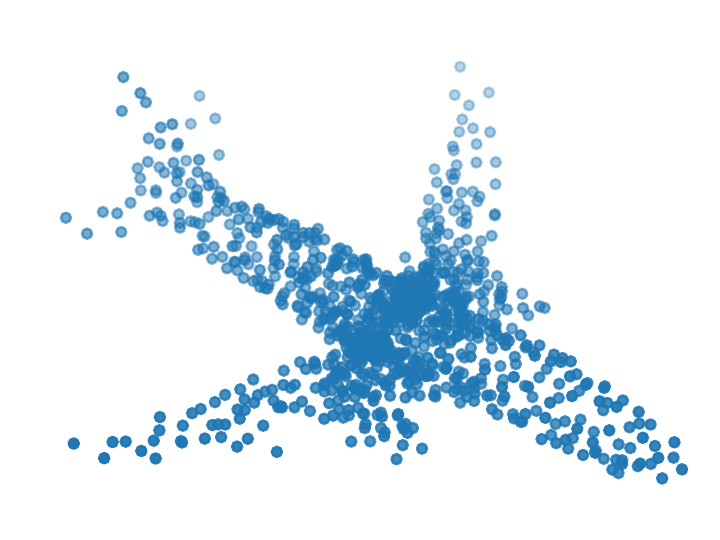}
      \caption{76 bits}
      \label{fig:rec/micro/4}
    \end{subsubfigure}%
    \hfill%
    \begin{subsubfigure}[b]{\subfigurehspace}
      \centering
      \includegraphics[width=\linewidth]{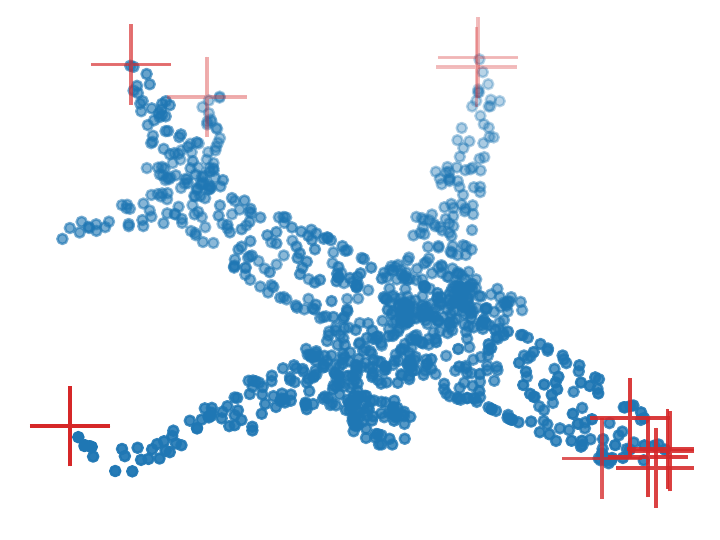}
      \includegraphics[width=\linewidth]{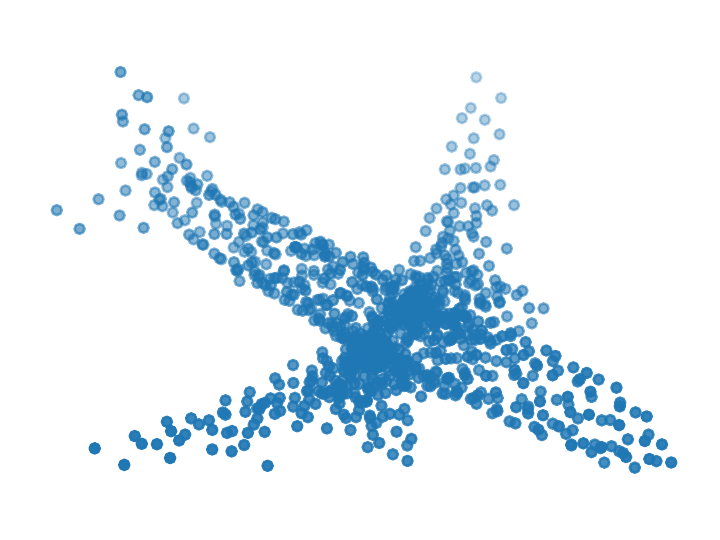}
      \caption{27 bits}
      \label{fig:rec/micro/3}
    \end{subsubfigure}%
    \hfill%
    \begin{subsubfigure}[b]{\subfigurehspace}
      \centering
      \includegraphics[width=\linewidth]{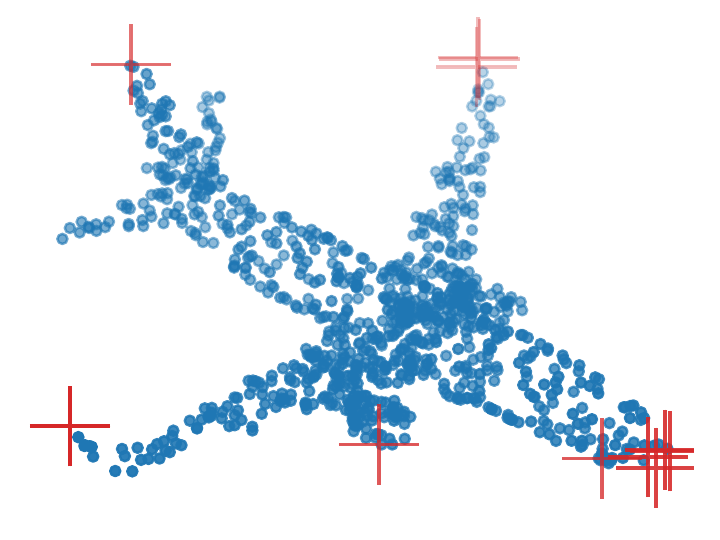}
      \includegraphics[width=\linewidth]{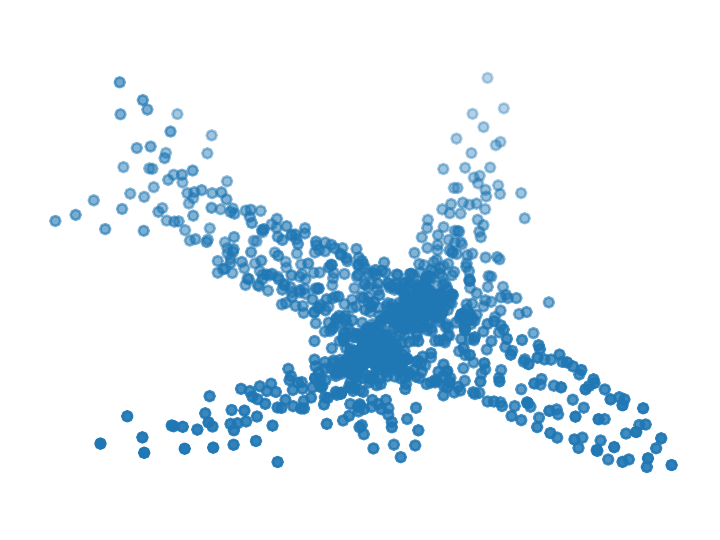}
      \caption{23 bits}
      \label{fig:rec/micro/2}
    \end{subsubfigure}%
    \hfill%
    \begin{subsubfigure}[b]{\subfigurehspace}
      \centering
      \includegraphics[width=\linewidth]{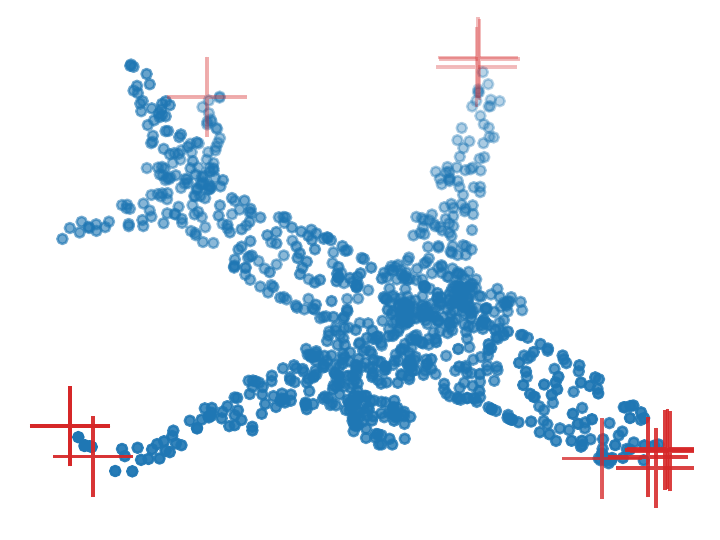}
      \includegraphics[width=\linewidth]{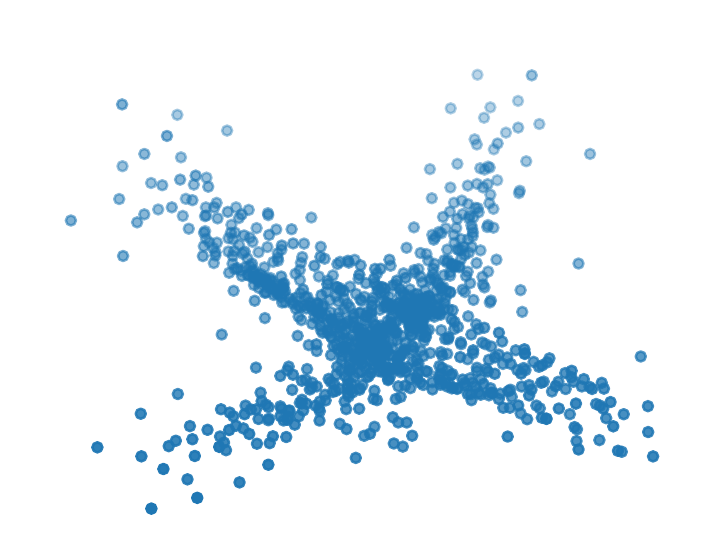}
      \caption{14 bits}
      \label{fig:rec/micro/1}
    \end{subsubfigure}%
    \caption{"micro" codec}
  \end{subfigure}%
  \caption{
    Reconstructions of a sample airplane 3D model from the ModelNet40 test set for various codecs and bitrates.
    For each reconstruction, its corresponding reference point cloud is marked with \emph{critical points} in red.
  }
  \label{fig:rec}
\end{figure}

In \cref{fig:rec}, we show various point clouds that have been reconstructed from the bitstreams generated by our proposed codecs.
For each codec, we include samples reconstructed from bitstreams compressed at different rates.
Above each reconstructed point cloud, we show the corresponding reference point cloud, with critical points marked in red.

\section{Discussion}
\label{sec:discussion}

For input point clouds containing $P = 1024$ points, our "full", "lite", and "micro" codec configurations achieve an accuracy of 80\% with as few as 30, 40, and 50 bits.
For comparison, $\log_2(40) \approx 5.3$ bits are required to losslessly encode uniformly distributed class labels of the 40 classes from ModelNet40.
Our codec comes surprisingly close to this theoretical lower bound, despite the fact that our architecture design omits the traditional MLP "classifier" within the encoder.
The same pointwise function is applied to all points, and the only operation that "mixes" information between the points is a pooling operation.
This suggests that our encoder should possess limited classification abilities, and yet it consumes only a few times more bits than the theoretical lower bound.

To put this into perspective, consider coding for image classification, which is one of the best developed areas in the field of coding for machines.
Current state-of-the-art (SOTA) approaches~\cite{matsubara2022wacv,duan2022pcs,Ahuja_2023_CVPR} for coding for image classification on ImageNet~\cite{ImageNet} require upwards of $0.01$ bits per pixel (bpp) to maintain a reasonable top-1 accuracy.
With the typical input image resolution of $224 \times 224$, this works out to be around $500$ bits.
However, the maximum classifier output entropy with $1000$ classes is only $\log_2(1000) \approx 10$ bits, which is several orders of magnitude lower.
Hence, the gap between the current SOTA and theoretical limits on coding for image classification is much higher than what is achieved by our proposed codec for point cloud classification.

To explore why our codec comes so close to the theoretical lower bound, we propose the following arguments.
Let $\boldvar{x}$ represent a possible input point cloud, and
let $\boldvar{\hat{y}} = (Q \circ g_a)(\boldvar{x})$ be its quantized transformed latent representation.
Applying the data processing inequality to the Markov chain
$\boldvar{x} \to \boldvar{x} \to \boldvar{\hat{y}}$,
we determine that
$I(\boldvar{x}; \boldvar{x}) \geq I(\boldvar{x}; \boldvar{\hat{y}})$.
Furthermore, since $Q \circ g_a$ is deterministic,
$H(\boldvar{\hat{y}} \mid \boldvar{x}) = 0$, and so
\begin{align*}
  H(\boldvar{x})
  = I(\boldvar{x}; \boldvar{x})
  \geq I(\boldvar{x}; \boldvar{\hat{y}})
  = H(\boldvar{\hat{y}}) - H(\boldvar{\hat{y}} \mid \boldvar{x})
  = H(\boldvar{\hat{y}}).
\end{align*}
This indicates theoretically that the quantized latent representation $\boldvar{\hat{y}}$ must on average be at least as compressible as the input point cloud $\boldvar{x}$ that it was derived from.

Since the critical point set $\boldvar{x}_C \subseteq \boldvar{x}$ produces the exact same $\boldvar{\hat{y}}$ as the original input point cloud $\boldvar{x}$, we may use the same arguments as above to argue that
\[ H(\boldvar{x}) \geq H(\boldvar{x}_C) \geq H(\boldvar{\hat{y}}). \]
This provides us with a potentially tighter bound.
In fact, as shown in \cref{fig:rec/micro/2}, much of the general shape of the shown sample point cloud can be reconstructed from only 23 bits of information.
Furthermore, since $|\boldvar{x}_C| \leq N$, there are only at most $32$ and $16$ distinct critical points for the "lite" and "micro" codecs, respectively.
This suggests part of the reason for why our proposed codec achieves such big gains in comparison to input compression.

\section{Conclusion}
\label{sec:conclusion}

In this paper, we proposed a new codec for point cloud classification.
Our experiments demonstrated that the "full" configuration of the codec achieves stellar rate-accuracy performance, far exceeding the performance of alternative methods.
We also presented "lite" and "micro" configurations of the codec whose encoders consume minimal computational resources, and yet achieve comparable gains in rate-accuracy performance.

Our work may be extended to other point cloud tasks, such as segmentation and object detection, or to more complex tasks involving larger models and larger point clouds from real-world datasets.
Our work also sets a good starting point for further research into approaches for scalable and multi-task point cloud compression.
We hope that our work will help towards achieving the end goal of more capable end devices.

\phantomsection

%
\bibliographystyle{IEEEtran}
\bibliography{references}


\end{document}